# Effect of Langmuir monolayer of bovine serum albumin protein on the morphology of calcium carbonate

Zhong-Hui Xue, Shu-Xi Dai, Bin-Bin Hu, Zu-Liang Du *

*Key Laboratory of Ministry of Education for Special Functional Materials, Henna University, Kaifeng 475004, PR China*

A B S T R A C T

Bovine serum albumin (BSA) Langmuir monolayer, as a model of biomineralization-associated proteins, was used to study its effect on regulated biomineralization of calcium carbonate. The effects of the BSA Langmuir monolayer and the concentration of the subphase solution on the nucleation and growth processes and morphology of the calcium carbonate crystal were investigated. The morphology and polymorphic phase of the resulting calcium carbonate crystals were characterized by scanning electron microscopy (SEM) and X-ray diffraction analysis (XRD). Moreover, the interaction mechanisms of the subphase solution with the BSA Langmuir monolayer were discussed. It was found that BSA Langmuir monolayer could be used as a template to successfully manipulate the polymorphic phase and crystal morphology of calcium carbonate and had obvious influence on the oriented crystallization and growth. The final morphology or aggregation mode of the calcite crystal was closely dependent on the concentration of calcium bicarbonate solution. It is expected that this research would help to better understand the mechanism of biomineralization by revealing the interactions between protein matrices and crystallization of calcium carbonate crystal.

## 1. Introduction

In nature, proteins are commonly used to direct and control the crystallization process of biominerals such as calcium carbonate with special orientation, texture, and morphology at ambient conditions [1]. Inspired by the formation of biominerals, a variety of mimetic methodologies have been designed and applied to investigate the possible formation mechanisms of natural biominerals [2,3]. Many soluble organic templates such as amino acids [4], polycarboxylic acids [5], synthetic peptides [6], and dendrimers [7] have been applied to control the mineralization process in vivo, following the idea of protein-directed biomineralization. In recent years, various proteins have been developed to control the mineralization process of calcium carbonate in solution, based on the fact that the biomineralized materials often contain proteins that are rich in glutamic or aspartic acid residues, and a variety of materials with novel microstructures have been successfully produced [8–12].

It is well known that a locally supersaturated solution is firstly generated by the organisms near the biomacromolecule surface when biominerals are deposited. Thus it could be feasible to simulate the biomineralization process by choosing suitable bioorganic surfaces to manipulate the supersaturated solutions of interest. Langmuir monolayer is an excellent candidate for this purpose: they are already in contact with water, they are ordered, and their structures are well-known and can be varied by changing pressure, temperature and other experimental parameters. Recently, the nucleation of inorganic crystals under Langmuir monolayer has been extensively studied [13–18]. Mann and coworkers have performed extensive studies of the nucleation of calcium carbonate [13,14], barium sulfate [15]. They found that preferentially nucleated crystals would be prepared under organic Langmuir monolayer with specific polymorphs and specific lattice orientations. Weiner and Traub have proposed an epitaxial model for aragonite formation in nacre with distinct [001] orientation [16,17]. Volkmer and coworkers demonstrated that charge density plays a key role in the oriented growth of $CaCO_3$ crystals under a monolayer of amphiphilic octa-acids [18]. However, the systematic understanding of additive controlled mineralization is still unclear. And few has been reported about the influence of protein on the precipitation of calcium carbonate under Langmuir monolayer, though some studies have been recently reported about the effect of proteins on the crystal growth of calcium carbonate in mixed solutions [19–21]. In the present work, the influence of bovine serum albumin Langmuir monolayer on the growth process and morphology evolution of calcium carbonate crystals at different subphase concentrations was investigated. We aim at exploiting a simple and facile method to preparing unusual structures via properly adjusting the subphase concentration. The possible biomineralization-associated formation mechanism of $CaCO_3$ crystals is also discussed. We are hoping to acquire more insights into the biomineralization mechanism and

* Corresponding author.
*E-mail address:* zld@henu.edu.cn (Z.-L. Du).

1 DTHKSEIAHRFKDLGEEHFKGLVLIAFSQYLQQCPFDEHVKLVNELTEFA
51 KTCVADESHAGCEKSLHTLFGDELCKVASLRETYGDMADCCEKEQPERNE
101 CFLSHKDDSPDLPKLKPDPNTLCDEFKADEKKFWGKYLYEIARRHPYFYA
151 PELLYANKYNGVFQECCQAEDKGACLLPKIETMREKVLTSSARQRLRCAS
201 IQKFGERALKAWSVARLSQKFPKAEFVEVTKLVTDLTKVHKECCHGDLLE
251 CADDRADLAKYICBBZBTISSKLKECKDPCLLEKSHCIAEVEKDAIPEDL
301 PPLTADFAEDKDVCKNYQEAKDAFLGSFLYEYSRRHPEYAVSVLLRLAKE
351 YEATLEECCAKDDPHACYTSVFDKLKHLVDEPQNLIKZBCBZFEKLGEYG
401 FQNALIVRYTRKVPQVSTPTLVEVSRSLGKVGTRCCTKPESERMPCTEDY
451 LSLILNRLCVLHEKTPVESKVTKCCTESLVNRRPCFSALTPDETYVPKAF
501 DEKLFTFHADICTLPDTEKQIKKQTALVELLKHKPKATEEQLKTVMENFV
551 AFVDKCCAADDKEACFAVEGPKLVVSTQTALA 582

**Fig. 1.** Primary sequence of BSA. This protein also contains Asp and Glu (97 residues, highlighted in yellow), Arg and Lys (82 residues, highlighted in magenta), as well as Gly (16 residues, highlighted in blue). (For interpretation of the references to color in this figure legend, the reader is referred to the web version of this article.)

provide practical guidance to the synthesis of new special functional materials.

## 2. Experiment

### *2.1. Materials*

Analytical grade $CaCO_3$ was obtained from Tianjin Institute of Biological Products (Tianjin, China). Analytical grade bovine serum albumin was commercially obtained (Sigma, USA; primary sequence shown in Fig. 1). Analytical grade amyl alcohol was commercially obtained from Tianjin Chemicals Co. Ltd., China. All solutions were prepared with MQ-water with a resistance of 18 MΩ $cm^{-1}$. The protein solutions were prepared with MQ-water, and 0.05% (v/v) amyl alcohol was added in all cases to improve the spreading process [22].

$Ca(HCO_3)_2$ solutions with different concentrations ($[Ca^{2+}]=2.5$, 5.0, and 7.5 mM) were prepared according to the procedures reported by Kitano [23]. The resulting supersaturated solution had a pH value of 7.2–7.4. The protein monolayer at the air–water interface was formed by spreading the BSA solution (0.1 mM) on the supersaturated calcium bicarbonate solutions. A 30-min lapse time was estimated to be sufficient for the protein monolayer to equilibrate before compression. A very low compression rate of ca. 3 mm/min was kept in the present research to obtain reproducible BSA isotherms [24]. All the experiment is carried out at constant temperature of 25 °C.

### *2.2. Crystallization of $CaCO_3$ under Langmuir monolayer of BSA*

The prepared supersaturated calcium bicarbonate solutions were poured into a Langmuir trough (KSV Minitrough), and the air–water interface was swept and aspirated before deposition of the surfactant solution. Pressure–area isotherms were recorded until the pre-set pressure was reached (15 mN $m^{-1}$). Each experiment was repeated three times under the same condition. The crystals as-grown in association with the BSA monolayer were obtained after crystallization for 12 h, by carefully horizontally pushing hydrophilic glass slides through the air–water interface to allow the direct deposition of the crystals (face growing into the solution) on the glass slide.

### *2.3. Characterization of crystal phase and morphology of $CaCO_3$*

The sizes and morphologies of the $CaCO_3$ crystals were characterized using a JSM-5600LV scanning electron microscope (SEM, JEOL Co. Ltd., Japan, operating at 20 kV). The glass slides with prepared crystals were mounted on copper sample stubs with a conducting carbon tape and sputter-coated with gold prior to viewing. The XRD measurements were carried out on an X'Pero Pro X-ray diffractometer (XRD, Philips Co. Ltd., Holland) at a scan rate of 0.080° $s^{-1}$, using Cu-K$\alpha_1$ radiation ($\lambda = 1.5406$ Å) operating at 40 kV and 40 mA as the excitation source.

## 3. Results and discussion

### *3.1. Surface pressure–area curve of bovine serum albumin*

Fig. 2 shows the surface pressure–area isotherms of the BSA protein spreading on pure water and supersaturated calcium bicarbonate solutions ($[Ca^{2+}]=2.5$, 5.0, and 7.5 mM). BSA can form a stable Langmuir monolayer on water surface as reported previously [24–26]. The pressure–area curves present that the BSA monolayers have similar surface pressure-isothermal shapes spreading on the pure water and supersaturated calcium bicarbonate solution. And BSA Langmuir monolayers on the supersaturated $Ca(HCO_3)_2$ solutions have a same collapse pressure of about 50 mN/m as that on pure water. Sanchez-Gutierrez et al. had studied a variety of transfer parameters for obtain the best conditions to get stable and compact BSA Langmuir–Blodgett films on water subphase [24]. They found that the transfer surface pressure near 15 mN/m is optimum because it guarantees a highly packed state and no segment readsorption of the molecule from the subphase [24]. We also chose the surface pressure of 15 mN/m to transfer BSA Langmuir monolayer on pure water and supersaturated calcium bicarbonate solution.

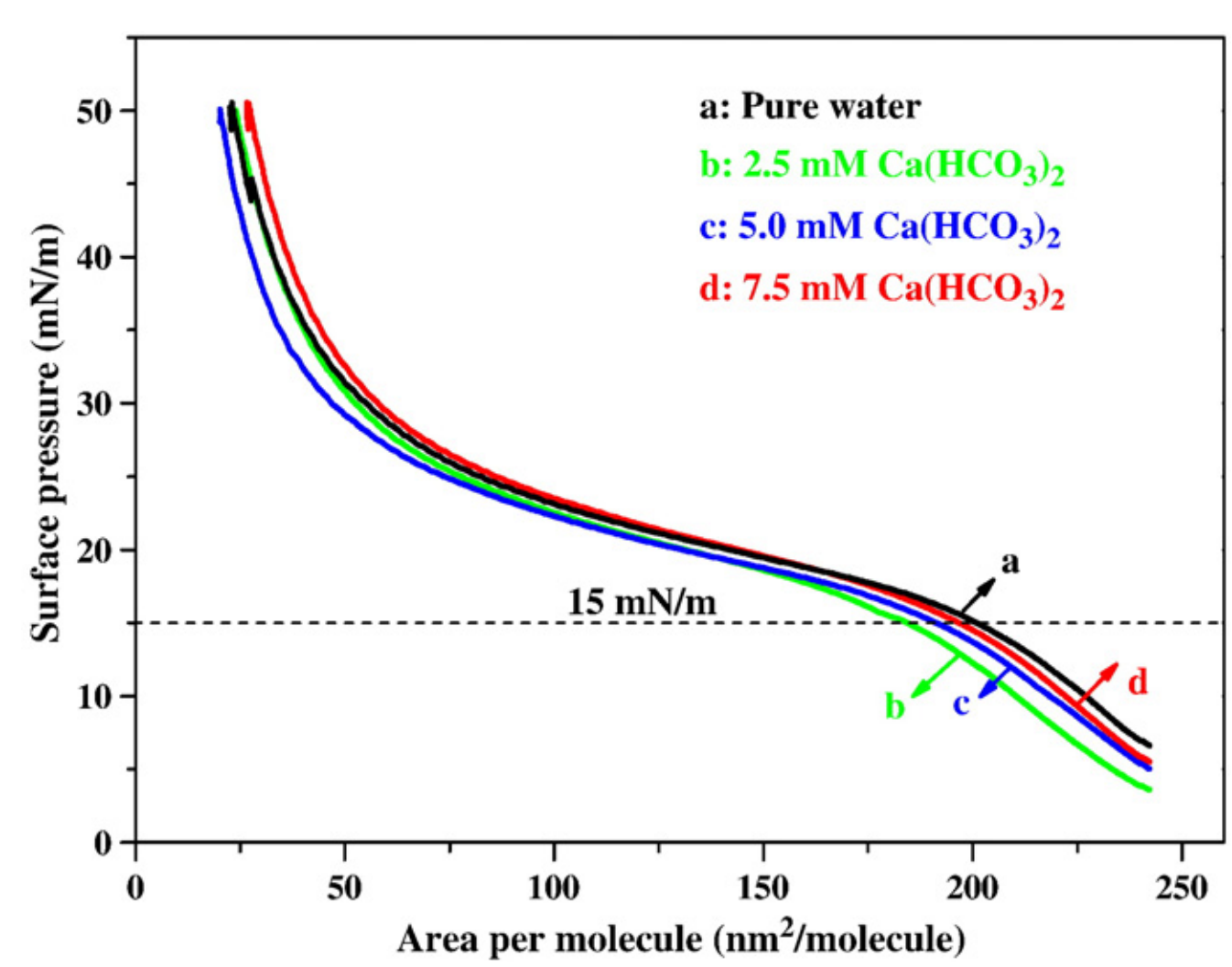

**Fig. 2.** The surface pressure–area isotherm of BSA spreading on the different subphase containing supersaturated calcium bicarbonate and pure water. (a) Pure water, (b) 2.5 mM, (c) 5.0 mM and (d) 7.5 mM.

Since BSA has an isoelectric point of 4.7, and supersaturated $Ca(HCO_3)_2$ solution has a pH value about 7.2–7.4, the BSA Langmuir monolayer on the supersaturated $Ca(HCO_3)_2$ solution can be negatively charged. There exists strong electrostatic attraction between the BSA molecules and $Ca^{2+}$ ions in the solution, which made the isotherms spread on supersaturated $Ca(HCO_3)_2$ solutions are more condensed than that on pure water. At the same time, there also exists electrostatic repulsion between the $Ca^{2+}$ ions adsorbed under BSA Langmuir monolayer. Form Fig. 2b to Fig. 2d, we could find the isotherms became expanded at the surface pressure of 15 mN/m with the increased concentration of calcium bicarbonate solution from 2.5 mM to 7.5 mM. The average molecular area ($A_{ama}$) on different supersaturated $Ca(HCO_3)_2$ solution is in the order of $A_{2.5ama}>A_{5.0ama}>A_{7.5ama}$. In this comparison, we expect the mobility of BSA to be similar within each monolayer, while the spatial distribution of the headgroups and charge density of BSA molecular at the interface are variable. This result suggests that the lower subphase concentration, the higher headgroups and charge density at the monolayer.

### 3.2. XRD patterns

Fig. 3 shows the XRD patterns of $CaCO_3$ crystals prepared at different subphase concentrations. The calcium carbonate formed at different subphase concentrations is composed of calcite phase with orientated growth along (104) plane. The calcite obtained at a subphase concentration of 2.5 mM shows a low intensity of diffraction peaks corresponding to (104) plane (Fig. 3a, JCPDS: No. 83-0578).The sample obtained at a subphase concentration of 5.0 mM presents increased intensity of diffraction peaks corresponding to (104) plane (Fig. 3b, JCPDS: No. 72-1650). With further increase in the subphase concentration of up to 7.5 mM, the diffraction peak of (104) plane becomes increasingly stronger and a secondary peak of (208) plane appears (Fig. 3c, JCPDS: No. 72-1651). The preferential interaction between the BSA protein Langmuir monolayer and the (104) face of calcite indicates that these crystal surface locations are energetically and/or stereochemically favored by the proteins.

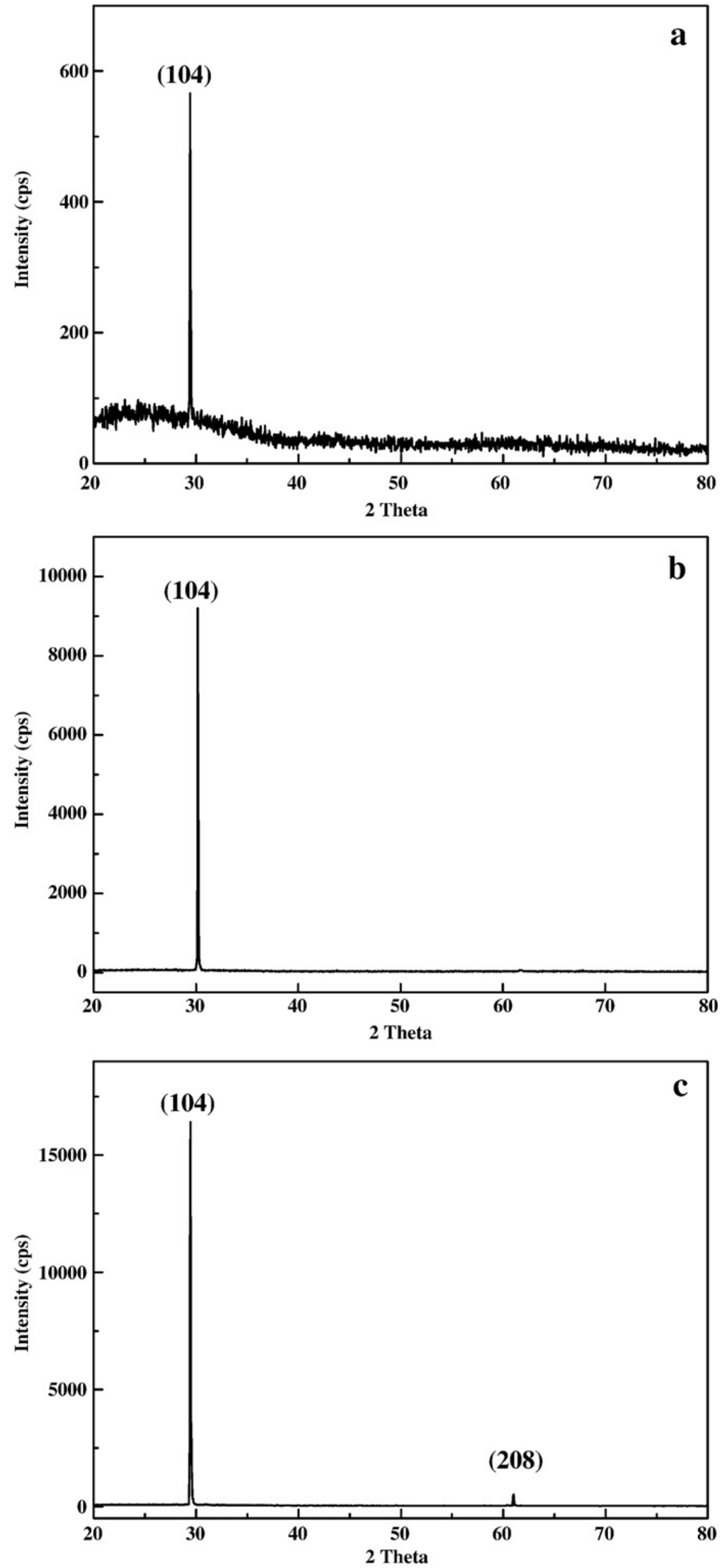

**Fig. 3.** XRD patterns of calcium carbonate formed at different subphase concentrations. (a) 2.5 mM, (b) 5.0 mM, (c) 7.5 mM.

### 3.3. SEM images

Fig. 4 presents the SEM morphologies of $CaCO_3$ crystals prepared at different subphase concentrations. The morphologies of the $CaCO_3$ crystals grown under the BSA Langmuir monolayer varied greatly with increasing concentration of the subphase solution. Disk-like calcite particles with an average diameter of 0.75 μm were produced at a subphase concentration of 2.5 mM (Fig. 4a, b). Bowknot-like calcite particles were formed at a subphase concentration of 5.0 mM (Fig. 4c, d). According to close SEM observation (Fig. 4d), and the bowknot-like calcite particles are composed of needle-like crystals. When the subphase concentration was further increased to 7.5 mM, the resulting calcite particles formed underneath the BSA Langmuir monolayer have hexagonal-shaped morphology with a size of 16–18 μm (Fig. 4e, f), while the boundaries of the hexagonal calcite are composed of small-sized plate-like crystals. SEM images show that the $CaCO_3$ samples under BSA monolayer consists of discrete crystals at low supersaturation. At the middle and high supersaturation, however, small crystals self-aggregate to form complex structure with different crystal morphologies. Thus it could be feasible to prepare $CaCO_3$ crystals with a variety of morphologies underneath the BSA Langmuir monolayer by properly adjusting the concentration of the calcium bicarbonate supersaturated solution.

Generally, the formation of the $CaCO_3$ crystals with different morphologies at different subphase concentrations corresponds to different nucleation and growth processes. In the present research, the constituent molecules or ions in the solution join into aggregates of two or more particles by collision to form dimers, trimers, tetramers, and so forth. Before the embryos reach a critical radius $r_c$, they are unstable even when a positive thermodynamic driving force $\Delta\mu$ is applied. A free energy barrier, so-called nucleation barrier, needs to be overcome in order to form nuclei of a critical radius $r_c$. After the process of overcoming the nucleation barrier, a second phase transition stage of growth begins. It has been reported that protein may recognize and bind to specific crystal faces, affecting nucleation, growth and morphology [27]. In this study, the protein macromolecules first formed a framework at the solution surface, and calcium ions were bound to the biopolymer chains via the carboxyl groups, resulting in a locally high concentration of calcium ions inside the framework, consequently inducing the nucleation. This process is similar to that occurring in the biomineralization process in vivo [28]. According to the nucleation and growth theory [29,30], the activation energy ($\Delta G_N$) must be overcome to form a new nucleus. $\Delta G_N$ can be expressed as:

$$\Delta G_N = 16\pi(\Delta G_1)^3 / 3(kT \ln S)^2$$

$$\Delta G_B = (kT \ln S)$$

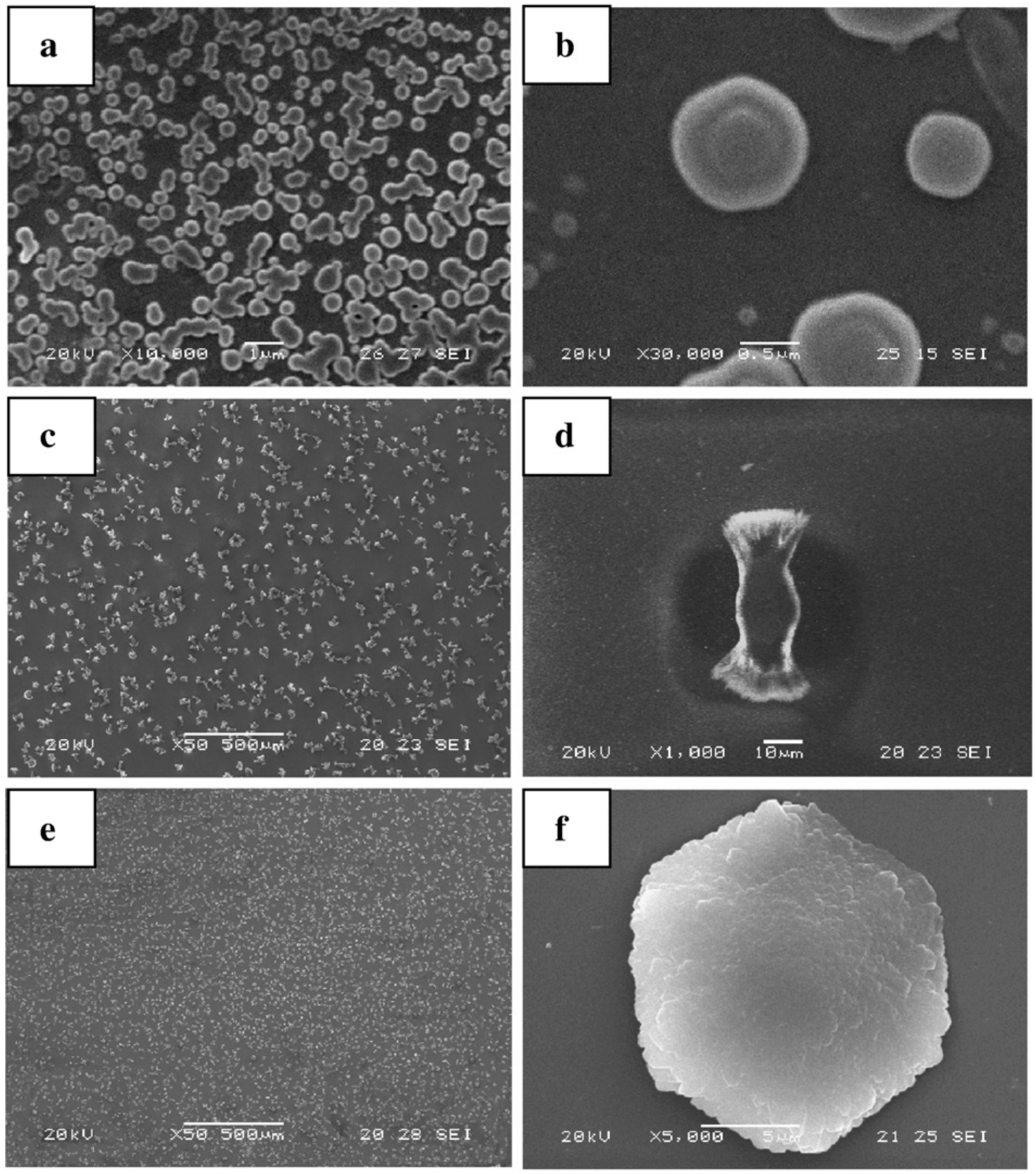

**Fig. 4.** SEM images of calcium carbonate formed at different subphase concentrations. (a and b) 2.5 mM, (c and d) 5.0 mM, (e and f) 7.5 mM.

Where $\Delta G_1$ is the surface energy needed to form the new interface and maintain the crystal growth, $\Delta G_B$ is the binding energy of crystal, *k* is Boltzmann constant, *T* is temperature, and *S* is the supersaturation degree of the solution. As seen in the above equations, the decrease of surface energy $\Delta G_1$ or the increase of *S* leads to decrease of the activation energy for crystal nucleation.

BSA molecule (582 AA; Fig. 1) has an isoelectric point (pI) of 4.7, lower than the pH (7.2–7.4) of subphase solution. Thus the surface of the BSA Langmuir monolayer should have surplus negative charges, implying that BSA could strongly attract $Ca^{2+}$ ions in the solution via electrostatic interactions, leading to an enrichment of local $Ca^{2+}$ concentration. In the present work, the concentration of $Ca^{2+}$ ions at the monolayer surface is generally independent of the concentration in the bulk [31], implying that it should be the BSA Langmuir monolayer that has decisive effect on the nucleation of $CaCO_3$. In addition to the putative BSA cation-binding by Asx residues, the presence of Lys and Arg residues may provide active sites for interaction with carbonate counterion. These acidic and basic residues, in conjunction with Ser and Glx residues, also provide active sites for hydrogen-bonding with water, and the presence of Gly residues generate considerable flexibility along the protein backbone. At the same time, owing to the dissolved $CO_2$ coming from the air–solution interface, the pH value at the interface would be reduced and the degree of supersaturation would be increased [32–33]. Moreover, the active nucleation sites offered by BSA for the growth of $CaCO_3$ particles could reduce the surface energy, leading to reduction of $\Delta G_N$, favoring the formation of the most stable high-energy phase.

Aside from the BSA Langmuir monolayer, some factors such as pH, temperature, and supersaturation degree also play key roles in controlling the growth and morphologies of $CaCO_3$ crystals. As one of the thermodynamic driving forces, the supersaturation degree is the most decisive parameter for crystallization process and has a tremendous influence on the mechanisms such as crystal nucleation, growth and aggregation, which will affect the morphology of crystals [30]. Therefore, the concentration of $Ca^{2+}$ ions in the aqueous bulk phase should have a vital effect on the crystal growth and morphology of $CaCO_3$, while the BSA Langmuir monolayer could function to modify the morphology of calcium carbonate via the interaction with $Ca^{2+}$ ions in the subphase solution.

To sum up, the morphological differences of the $CaCO_3$ crystals formed at different subphase concentrations could be attributed to the variation in the subphase concentration and the interactions between the BSA Langmuir monolayer and $Ca^{2+}$ ions in the subphase solution. Further work is underway to reveal the role of the BSA Langmuir monolayer and subphase concentration in the nucleation and growth processes of the calcium carbonate at more details.

## 4. Conclusions

Pure calcite particles with different morphologies were successfully prepared through biomimetic routes from the supersaturated calcium bicarbonate solution underneath BSA Langmuir monolayer as a model of biomineralization-associated proteins. The BSA Langmuir monolayer showed obvious influence on the orientation crystallization, growth processes and hence the morphology of calcium carbonate crystal, while the concentration of calcium bicarbonate solution obviously influenced the final morphology or aggregation mode of calcite crystals. This research could provide important information about the interaction between protein matrix and crystallization of calcium carbonate, which is useful to understand the mechanism of biomineralization.

## Acknowledgements

This work was supported by Natural Science Foundation of China (No. 20371015), Program for New Century Excellent Talents in University (No. NCET-04-0653).